# THz-Frequency Modulation of the Hubbard U in an Organic Mott Insulator

R. Singla[1], G. Cotugno[1,2], S. Kaiser[1,7,8], M. Först[1], M. Mitrano[1], H. Y. Liu[1], A. Cartella[1], C. Manzoni[1,4], H. Okamoto[5], T. Hasegawa[6], S.R. Clark[2,9], D. Jaksch[2,3], A. Cavalleri[1,2]

[1]Max Planck Institute for the Structure and Dynamics of Matter, Luruper Chaussee 149, 22761 Hamburg, Germany

[2]Department of Physics, Oxford University, Clarendon Laboratory, Parks Road, OX1 3PU Oxford, United Kingdom

[3]Centre for Quantum Technologies, National University of Singapore, 3 Science Drive 2, Singapore 117543, Singapore

[4]IFN-CNR, Dipartimento di Fisica-Politecnico di Milano, Milan, Italy

[5]Department of Advanced Material Science, University of Tokyo, Chiba 277-8561, Japan

[6]National Institute of Advanced Industrial Science and Technology (AIST), Tsukuba 305-8562, Japan

[7]Max Planck Institute for Solid State Research, Heisenbergstraße 1, 70569 Stuttgart, Germany

[8]4th Physics Institute, University of Stuttgart, Pfaffenwaldring 57, 70550 Stuttgart, Germany

[9]Department of Physics, University of Bath, Claverton Down, Bath BA2 7AY, United Kingdom

Email: rashmi.singla@mpsd.mpg.de, stefan.kaiser@mpsd.mpg.de, andrea.cavalleri@mpsd.mpg.de

**Abstract:** We use midinfrared pulses with stable carrier-envelope phase offset to drive molecular vibrations in the charge transfer salt ET-F2TCNQ, a prototypical one-dimensional Mott insulator. We find that the Mott gap, which is probed resonantly with 10 fs laser pulses, oscillates with the pump field. This observation reveals that molecular excitations can coherently perturb the electronic on-site interactions (Hubbard U) by changing the local orbital wave function. The gap oscillates at twice the frequency of the vibrational mode, indicating that the molecular distortions couple quadratically to the local charge density.

Nonlinear phononics, the coherent excitation of anharmonically coupled collective lattice modes, can be used to create transient crystal structures with new electronic properties [1]. Examples of such nonlinear phonon control are ultrafast insulator-metal transitions [2], melting of magnetism [3] and light induced superconductivity [4,5,6]. Nonlinear phononics has been realized and understood in the limit of cubic anharmonic coupling [7,8], which results in rectification of an excited lattice oscillation and in the net displacement of the atomic positions along a second vibrational mode. Change in the electronic properties results then from the altered bond angles and nearest neighbor atomic distances [9], which perturb hopping amplitudes and exchange interactions. However, local parameters like Mott correlations, of importance in many complex materials, are not modulated by nonlinear phononics.

In this paper, we show that the onsite Coulomb integral can be modulated in molecular solids by driving local molecular degrees of freedom to large amplitudes [10,11]. The excitation of local modes at mid-infrared frequencies is different from the case of nonlinear phononics in that the molecular orbital and concomitantly the onsite charge density are controlled [12,13], with each site maintained in its electronic ground state.

A Mott insulator is a half-filled solid in which electrons are made immobile by their mutual Coulomb repulsion. This physics is often described by the extended Hubbard Hamiltonian

$$\hat{H} = -t \sum_{j\sigma} (\hat{c}_{j\sigma}^+ \hat{c}_{j+1\sigma} + h.c.) + V \sum_j \hat{n}_j \hat{n}_{j+1} + U \sum_j \hat{n}_{j\uparrow} \hat{n}_{j\downarrow} \qquad (1)$$

where $U$ and $V$ represent on-site and nearest neighbor site Coulomb repulsions, and $t$ denotes the tight binding hopping amplitude [14,15]. In equation (1), $\hat{c}_{j\sigma}^\dagger (\hat{c}_{j\sigma})$ is the creation (annihilation) operator for an electron at site $j$ with spin $\sigma$, while $\hat{n}_{j\sigma}$ is the associated number operator and $\hat{n}_j = \hat{n}_{j\uparrow} + \hat{n}_{j\downarrow}$. The key features of the Hubbard model are well reproduced in some molecular solids, and, in one dimension, by the charge transfer salt Bis-(ethylendithyo)-tetrathiafulvalene-difluorotetracyano-quinodimethane (ET–F$_2$TCNQ). The crystal structure of this material is shown in Fig. 1a. One-dimensional Mott physics is observed along the crystallographic *a*-axis, where a half filled chain of ET molecules is characterized by small intersite tunneling amplitude ($t \sim 40$ meV) and large Coulomb repulsion ($V \sim 120$ meV and $U \sim 840$ meV). Hence, although the system is fractionally filled, a large correlation gap is found. As shown in Fig. 1b, the reflectivity spectrum of ET–F$_2$TCNQ exhibits a sharp charge transfer (CT) band centered at 5500 cm$^{-1}$ [16]. Note, the small electron-lattice interaction in ET-F$_2$TCNQ [17] is insufficient to drive a Peierls distortion, so that Mott physics is retained down to very low temperatures. Moreover, since our experiments are performed at room temperature, far above the 30 K Neel temperature [16], and $U \gg k_BT \gg t^2/U$ no long-range Neel spin-order is present.

In this and other charge transfer crystals, one finds collective phonon modes only at very low frequencies, well separated from higher-frequency localized molecular vibrations, which are observed in the mid-infrared. We here consider the physical situation discussed schematically in Fig. 2. A single localized vibration of the ET molecule is resonantly driven along the electric field of a mid-infrared optical pulse. This vibrational excitation leads to a time-dependent deformation of the valence orbital wavefunction and variation of local charge densities at twice the frequency of driving field. This in turn modulates the Hamiltonian parameters introduced in Eq. (1). To account for this physics, extra terms are added whose form is justified as follows: First, we assume that the classical vibrational mode coordinate $q_j$ only couples to the local charge density, and neglect the coupling to other lattice modes [18]. In general this gives additional terms of the form $\hat{n}_j f(q_j) + \hat{n}_{j\uparrow}\hat{n}_{j\downarrow}g(q_j)$, where $f(q_j)$ and $g(q_j)$ are two functions of the local mode coordinate that are not known a priori. Expanding the functions $f$ and $g$ as a series we obtain

$$\hat{H}_{\text{e-vib}} = \sum_j \hat{n}_j (A_1 q_j + A_2 q_j^2 + \cdots) + \sum_j \hat{n}_{j\uparrow}\hat{n}_{j\downarrow}(B_1 q_j + B_2 q_j^2 + \cdots). \quad (2)$$

Since the molecule is centrosymmetric and the vibrational mode is of odd symmetry, the terms linear in $q_j$ [19,20], which dominate the coupling for all even modes, vanish ($A_1 = B_1 = 0$). As the vibrational mode's natural frequency is $\Omega_{\text{IR}}$, every molecule is coherently driven by the laser pulse with its coordinate in time $\tau$ described as $q_j(\tau) = q_{\text{IR}}(\tau) = C \sin(\Omega_{\text{IR}} \tau)$, where $C$ is the driving amplitude. This implies that the $A_2$ term $\propto q_{\text{IR}}^2(\tau) \sum_j \hat{n}_j$ couples to the total density, resulting in an irrelevant global phase shift. We are therefore left with a quadratic coupling to the onsite interaction of the form $\hat{H}_{\text{e-vib}} = B_2 q_{\text{IR}}^2(\tau) \hat{n}_{j\uparrow}\hat{n}_{j\downarrow} = \frac{C}{2} B_2 [1 - \cos(2\Omega_{\text{IR}} \tau)]\hat{n}_{j\uparrow}\hat{n}_{j\downarrow}$. Importantly, the coefficient $B_2 < 0$ because the vibration will in general cause the valence orbital to spatially expand.

Thus, two effects are expected as long as an odd molecular mode is driven: a time-averaged reduction of the onsite repulsion together with its modulation at $2\Omega_{\text{IR}}$.

These predictions can be tested only by exciting vibrational oscillations by mid-infrared pump pulses with stable carrier-envelope phase (CEP), in which the temporal offset between the electric field and the intensity envelope is constant and locked over consecutive laser shots. Hence, every laser pulse drives the molecule with identical phase and allows measurements performed with sub-cycle resolution over many pump pulses. The evolution of these oscillations is then monitored by a delayed probe pulse. The time resolution of this experiment is given by the probe-pulse duration, which should be of the order of one half of the vibrational period to resolve effects at $\Omega_{\text{IR}}$ and one quarter to observe oscillations at $2\Omega_{\text{IR}}$. In the experiments on ET-F$_2$TCNQ discussed here, a 10 μm vibrational mode with a period of 33 fs is excited, requiring probe pulses of approximately 8 fs duration, tuned to the 700 meV charge transfer resonance. Note that this pulse duration corresponds to less than two optical cycles at the corresponding probe wavelength of 1.7 μm.

In order to generate the required ultrashort probe pulses, the following system was developed. Mid-infrared optical pulses with locked carrier envelope phase were obtained by difference frequency generation from two near-infrared (NIR) optical parametric amplifiers (OPA) driven by an amplifed Ti:sapphire laser operating at 800 nm wavelength and at 1 kHz repetition rate. The two OPAs were seeded by replicas of a single white-light continuum to preserve their relative phase; difference frequency mixing between the two NIR amplified sources occurring in a GaSe crystal generated passively CEP-stable 130 fs pulses tunable in the mid-infrared region [21]. To compensate for slow thermal drifts, these pulses were actively stabilized in phase [22] and used to drive the infrared-active molecular mode of ET–$F_2$TCNQ at 10 μm wavelength (30 THz frequency), perpendicular to the *a*-axis (see Fig. 3a). The system was probed by 700 meV pulses derived from a third, near-infrared OPA pumped by another portion of the 800 nm Ti:sapphire laser source. The broadband probe pulses, an order of magnitude weaker than the pump pulse, covered the spectral region between 570-980 meV (4600-7900 $cm^{-1}$) as shown in Fig. 3a; their duration was compressed to approximately 10 fs, close to the transform limit, using a deformable mirror in a grating-based 4-f pulse-shaper [23]. Figure 3b shows the measured phase-locked electric field of the pump pulse and the intensity envelope of the probe pulse. The probe beam was polarized along the *a*-axis, evincing the one-dimensional physics along ET chain. The experiments were performed at room temperature.

Figure 3c reports the spectrally integrated time-resolved reflectivity changes ∆R/R in ET-$F_2$TCNQ for a pump fluence of 0.9 mJ/$cm^2$, obtained by detecting the energy of the reflected probe pulses with a photo-diode using lock-in technique. The reflectivity was observed to reduce by 4 % during the pump pulse, recovering in 45 fs along the trailing edge of the pump pulse and followed by a slower exponential decay of 500 fs. The fast decay is attributed to the direct relaxation of the driven vibrational mode. The slower response is attributed to lattice thermalization. Crucially, fast oscillations at approximately 70 THz were observed, with an amplitude of 0.7 %. Note that these oscillations are convolved with the 10 fs duration of the probe pulse, longer than their 8 fs half-period, thus reducing their amplitude. In the same figure, we also show the response after deconvolution, now exhibiting oscillations of about the same amplitude as the overall reduced reflectivity (grey trace in Fig. 3c).

The microscopic origin of these dynamics appears to follow qualitatively the response discussed above for quadratic coupling of the type $\hat{H}_{e-vib} = \frac{C}{2}B_2[1 - \cos(2\Omega_{IR}\tau)]\hat{n}_{j\uparrow}\hat{n}_{j\downarrow}$. For a 1D system, any change in filling will bring the system into metallic state [24]. Since no metallic Drude response is observed in the THz regime [11], we can rule out a transient deviation from the average band filling as origin of the experimentally observed oscillations. Further insight was possible by performing a spectrally-resolved pump-probe measurement using a gated spectrometer for the detection of the probe. The time-dependent reflectivity spectrum is shown in Fig. 4a, evidencing a reduction in the total reflectivity (from a peak value of 0.4 to 0.33), a red shift of the reflectivity peak (by ~70 cm$^{-1}$), and oscillations with the same 70 THz frequency detected in the spectrally integrated measurements of Fig. 3c. The observed reflectivity change in Fig. 4a is consistent with the spectrally integrated data in Fig. 3c, when integrating the spectral changes over the weighted probe spectral content. From this reflectivity spectrum, the optical conductivity was calculated at each time delay by Kramers-Kronig (KK) consistent variational dielectric function fit [25]. The results are reported in Fig. 4b, clearly showing oscillations of the conductivity spectrum.

We next analyze the measured response by fitting these transient conductivity spectra. As discussed in Refs. [26,27,28], the charge transfer band of ET-F$_2$TCNQ in equilibrium is described by two dominant contributions to the optical conductivity (see Fig. 4c). The rather strong peak at lower frequency is related to holon-doublon (HD) pair excitations, that is, the transfer of charges between neighboring ET sites. As this holon-doublon exciton is bound by the intersite Coulomb correlation $V$, its energy is centered at $U - V - 4t^2/V$ [28]. A second weaker contribution to the conductivity spectrum results from the excitation of delocalized holon-doublon pairs not bound by the intersite Coulomb energy $V$ [15], which form a particle-hole (PH) continuum described by a band of width $8t$ centered at $U$ [27].

Figure 4c reports fit results of the optical conductivity, in which individually normalized contributions from bound HD pairs and and PH continuum are displayed in interval of ¼ of time period of the pump electric field. We find that the peak positions of both HD and PH oscillate at 70 THz frequency, synchronized with the molecular mode. We extract $U$ and $V$, fitting the optical conductivity along the lines described in Ref. [27], as a function of pump-probe delay time. Assuming that the hopping amplitude does not oscillate with the driven local vibrational motion [29], this procedure yields the effective correlation term $U/t$ and $V/t$. Crucially, the overall reduction of $U/t$ is superimposed by a response at frequency $2\Omega_{IR}$, consistent with a quadratic coupling of $U$ to the vibrational degree of freedom (see Fig. 5a). $V/t$ also shows a prominent reduction contributing to the transient changes (Fig. 5c) but does not show oscillatory behaviour. Note that as for the spectrally integrated measurements the spectral shifts (35 cm$^{-1}$ (4 meV) and 93 cm$^{-1}$ (12 meV) wavenumbers for HD and PH features, respectively) and the corresponding estimate of the $U/t$ modulations are strongly underestimated, because our probe beam is longer than the limit of 8 fs. No attempt to deconvolve this resolution factor was made for the $U/t$ parameter, although the response is likely to be one order of magnitude larger than estimated here. Note also that the absolute changes induced by the pump in $U$ and $V$ are of comparable magnitude, 7 meV and 14 meV respectively. Finally, the observation that the 70 THz oscillations are at slightly higher frequency than twice the excitation frequency (60 THz). This effect is likely related to the precision of our measurement. However, we cannot exclude that second order effects may be taking place, beyond the considerations discussed in Eq. (2). No attempt was made to analyze these possible effects, which may involve for example a change in frequency of the vibrations by fourth order coupling, potentially squeezing the vibration [30,31].

Additional theoretical analysis further substantiates the coherent modulation of the Hubbard parameter. An effective model describing electronic excitations above the insulating ground state in the strong-coupling limit was used (see supplemental material [32]), allowing for a numerical computation of the optical conductivity from the non-equilibrium two-time current-current correlation function $\chi_{JJ}(\tau,\tau') = \langle 0|\hat{\mathcal{U}}^\dagger(\tau+\tau')\hat{J}\hat{\mathcal{U}}(\tau+\tau')\hat{\mathcal{U}}^\dagger(\tau')\hat{J}\hat{\mathcal{U}}(\tau')|0\rangle$ [33,34,35]. In this expression $\hat{\mathcal{U}}(\tau)$ is the time-evolution operator of the system up to time $\tau$ including time dependence caused by the driving and $\hat{J}$ is the current operator. We assumed that $U$ varies in time as a Gaussian pulse envelope superimposed with the $q_{IR}^2(\tau)$ oscillation as $U(\tau) = U\{1 - A_U e^{-(\tau-\tau_P)^2/T_P^2}[1 - P_0 q_{IR}^2(\tau)]\}$, where $\tau_P$ and $T_P$ are the center and width of the pulse taken from the experiment, and $A_U$ quantifies the overall reduction of $U$ seen from the equilibrium fitting in Fig. 5a. Only the parameter $P_0$, giving the amplitude of the oscillations around the envelope, was fitted, giving a profile shown in Fig. 5b. We assumed that $V$ varies only with the envelope (Fig. 5d). The computed $\chi_{JJ}(\tau,\tau')$ was then transformed into the simulated reflectivity and simulated conductivity shown in Fig. 5e and 5f and is seen to give an excellent reproduction of measured reflectivity and conductivity shown in Fig. 4a and 4b. In order to rule out a modulation of the nearest neighbor interaction we repeated these calculations also under the assumption of an oscillatory *V/t* ratio. The calculated optical properties (reported in Fig. 3 of supplementary) differ significantly from the experimental observations, further supporting the claim that the onsite *U* modulation is the dominant effect behind the oscillatory behaviour.

In summary, the time-dependent reflectivity of the molecular Mott insulator ET-F$_2$TCNQ shows a coherent modulation of the correlation gap following phase-stable vibrational excitation. The nonlinear characteristic of the coupling is evidenced by the response, which occurs at approximately twice the frequency of the driving pulse. Fitting of the time dependent optical conductivity reveals that the onsite Hubbard-$U$ is being directly modulated by the excitation of the local molecular vibration. This is in contrast to what was discussed to date for the excitation of collective phonon modes. Vibrational control may be combined with mid-infrared pulse shaping techniques [36], well established in the visible spectral range [37, 38], and open up new avenues to coherently control interactions in a many-body system, a task to date only possible with cold-atoms optical lattices via the Feshbach resonance [35].

We acknowledge financial support the European Research Council under the European Union's Seventh Framework Programme (FP7/2007-2013)/ERC Grant Agreement No. 319286 Q-MAC.

**FIGURES**

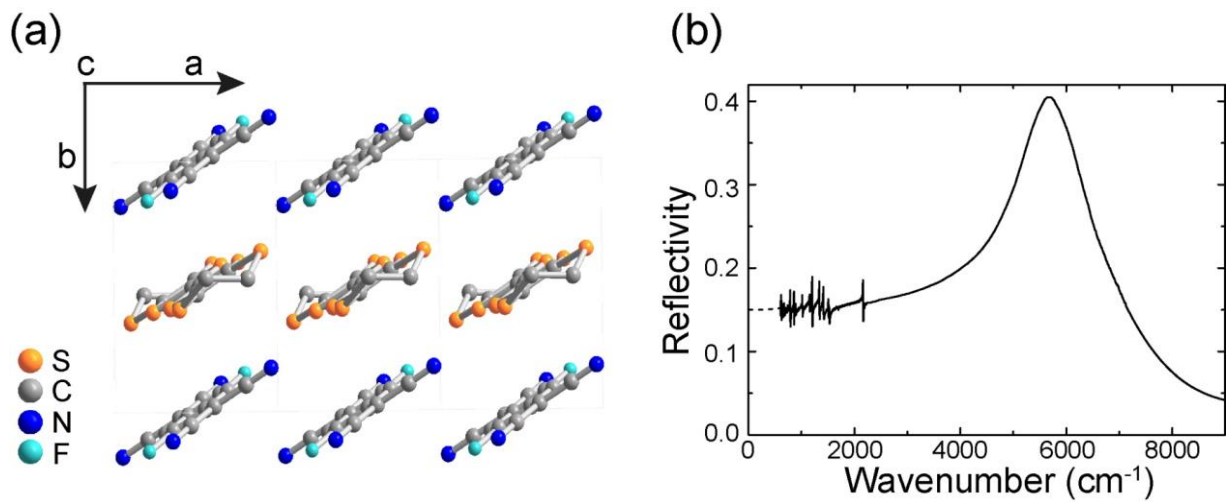

Figure 1: (a) The crystal structure of one-dimensional organic salt ET–$F_2$TCNQ (yellow=ET, blue=$F_2$TCNQ).

(b) Static reflectivity of the compound with electric field parallel to crystallographic *a*-axis (along ET chain).

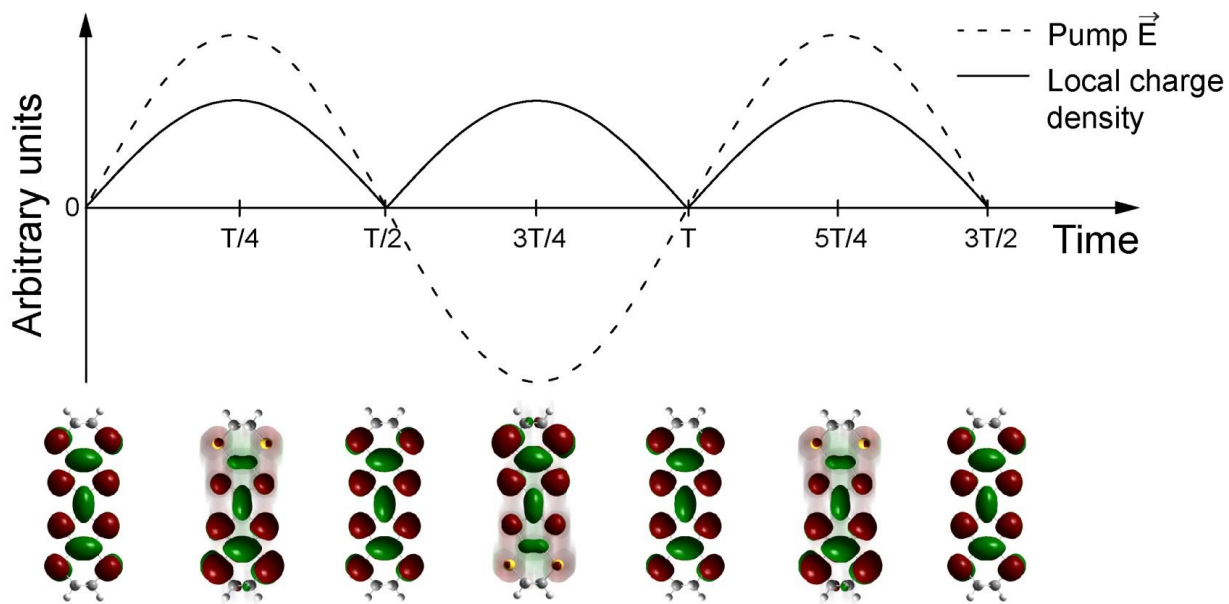

Figure 2: (a) Temporal evolution of the pump electric field (dashed black line, together with the resultant change in the local charge density (solid black line) and the corresponding orbital motion of vibrationally excited ET molecule over time.

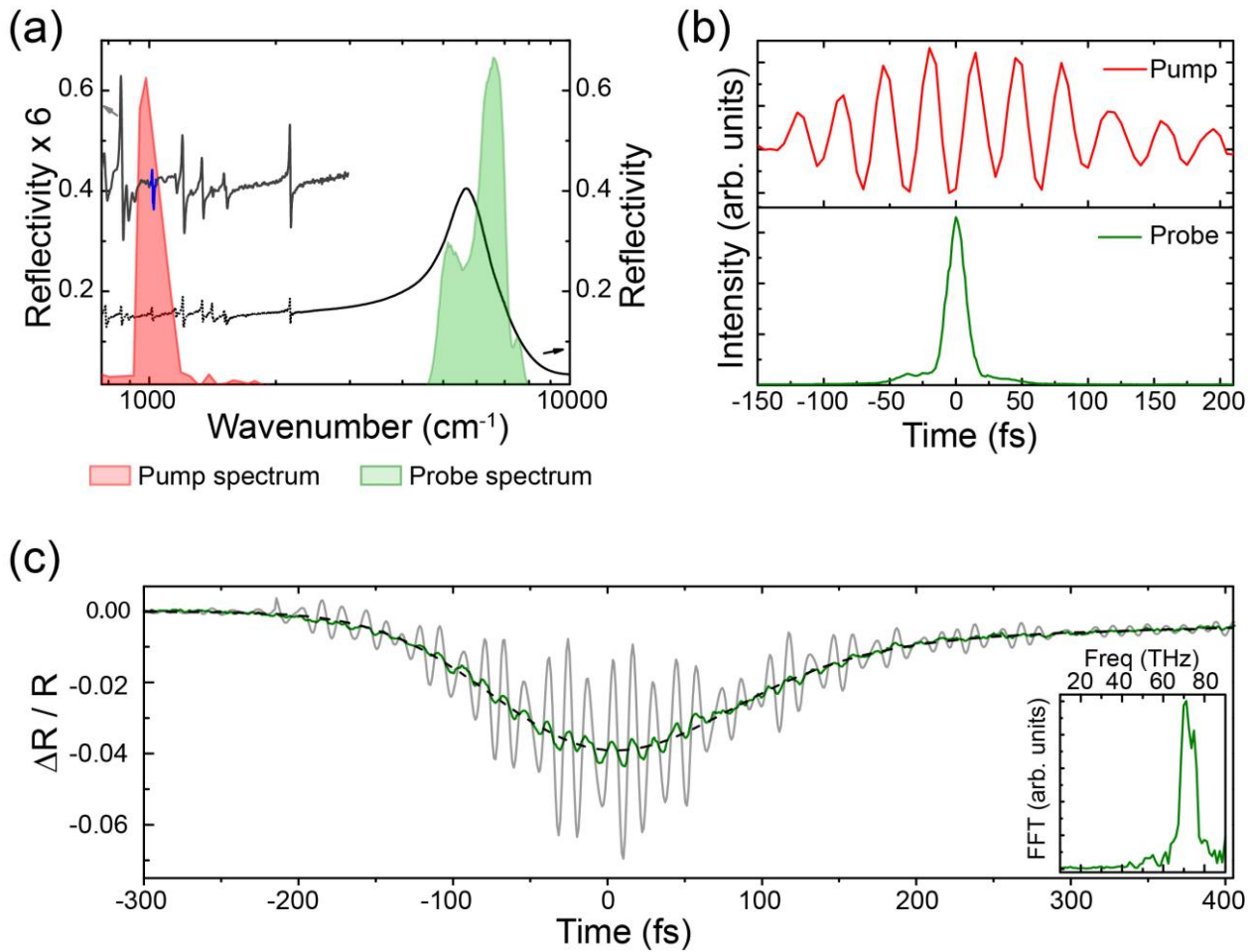

Figure 3: (a) (Red) Spectrum of the mid-infrared excitation pulses centered at 1000 cm$^{-1}$ (10 μm wavelength) resonant to the ET molecular vibration mode (blue). The sample reflectivity in direction perpendicular to the *a*-axis is shown in grey. (Green) The spectrum of the near-infrared probe beam covers the charge transfer band along a-axis. (b) (upper panel) Phase locked electric field of the mid-infrared pump pulse measured via electro-optic sampling, (lower panel) Intensity profile of NIR probe pulse. (c) (Green) Spectrally integrated time dependent reflectivity changes in ET-F$_2$TCNQ at a pump fluence of 0.9 mJ/cm$^2$ at room temperature, together with a double exponential fit of the form $A \, \mathrm{erf}(t/t_d + 1) \sum_{i=1,2} A_i \, e^{-t/\tau_i}$ in dashed black. The grey solid line shows the deconvolved reflectivity changes, and the inset shows a Fourier transformation of the measured oscillations, peaking at 70 THz.

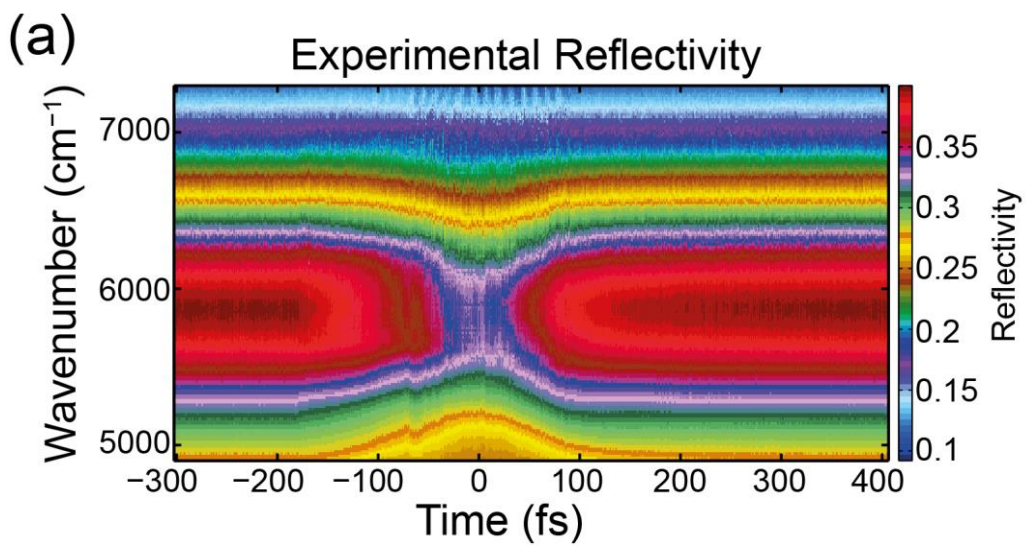

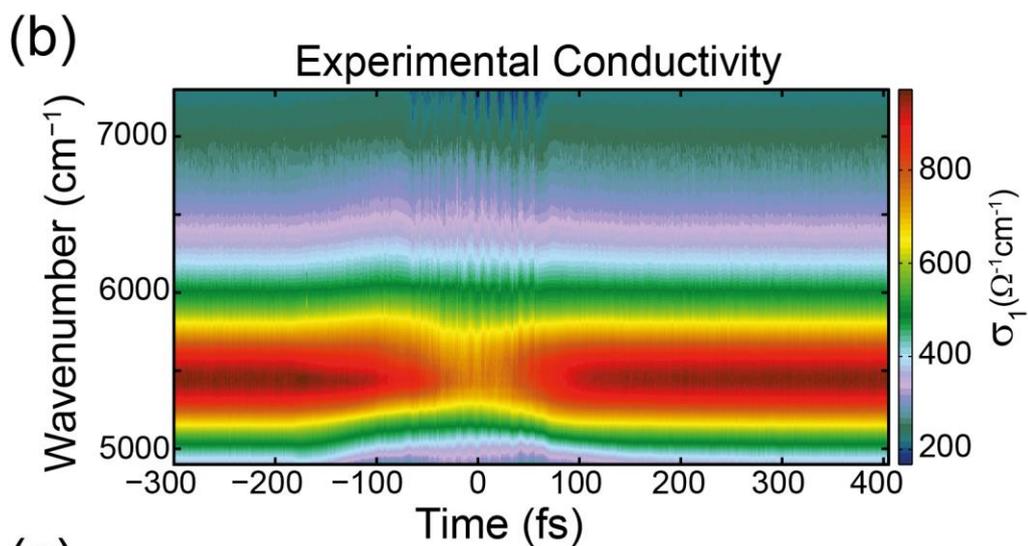

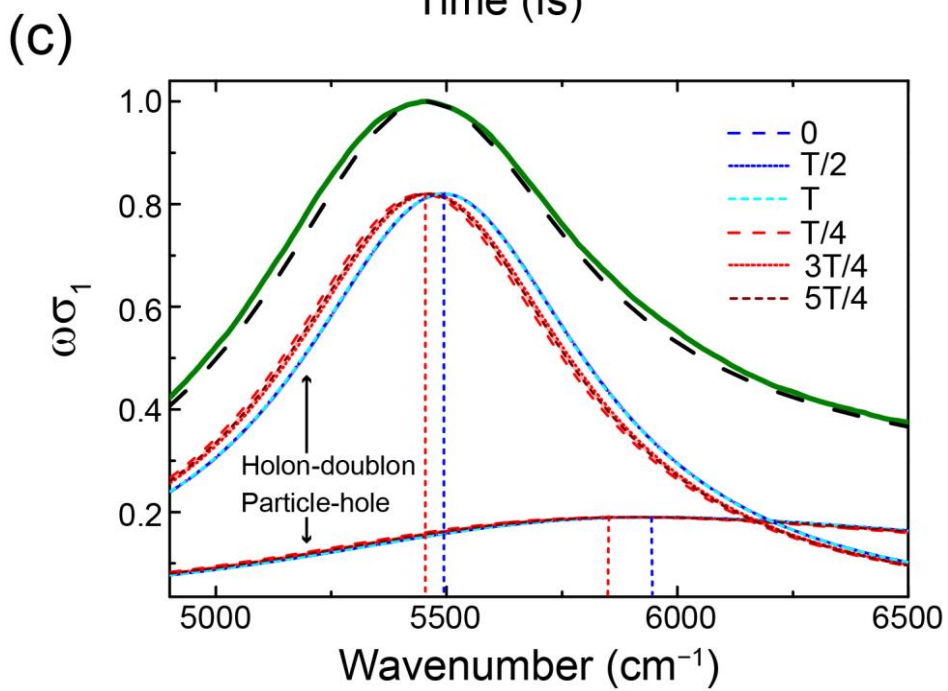

Figure 4: (a) Frequency-resolved reflectivity as function of pump-probe delay time. (b) Corresponding optical conductivity extracted from reflectivity using KK variational dielectric fit function. (c) Normalized equilibrium optical conductivity (green) of ET-F$_2$TCNQ at room temperature with model fit (dashed black). Below are the fitted contributions from HD and PH continuum, individually normalized at various time along the pump electric field cycle of time period T.

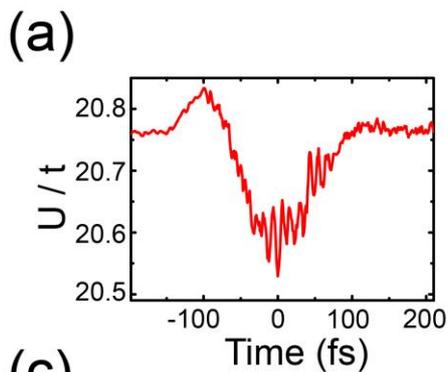
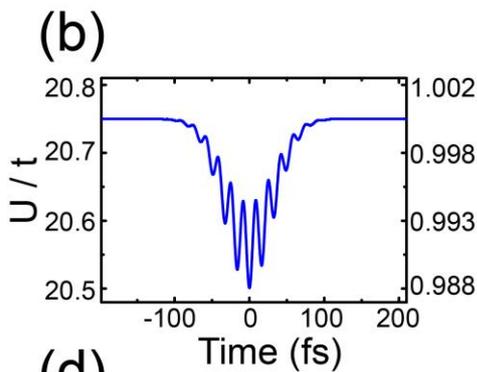
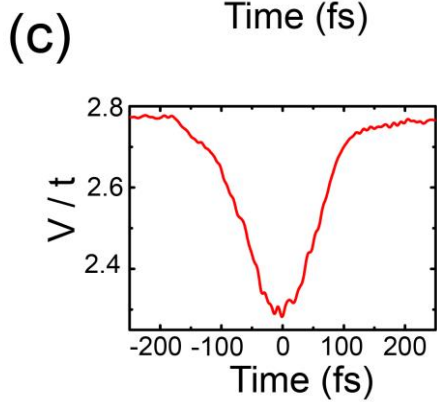
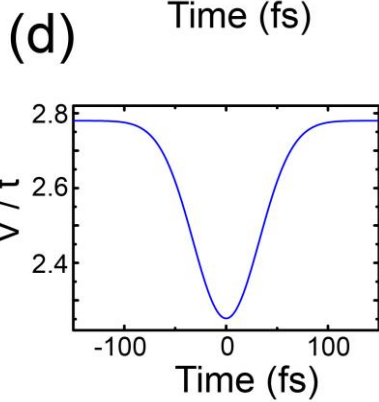
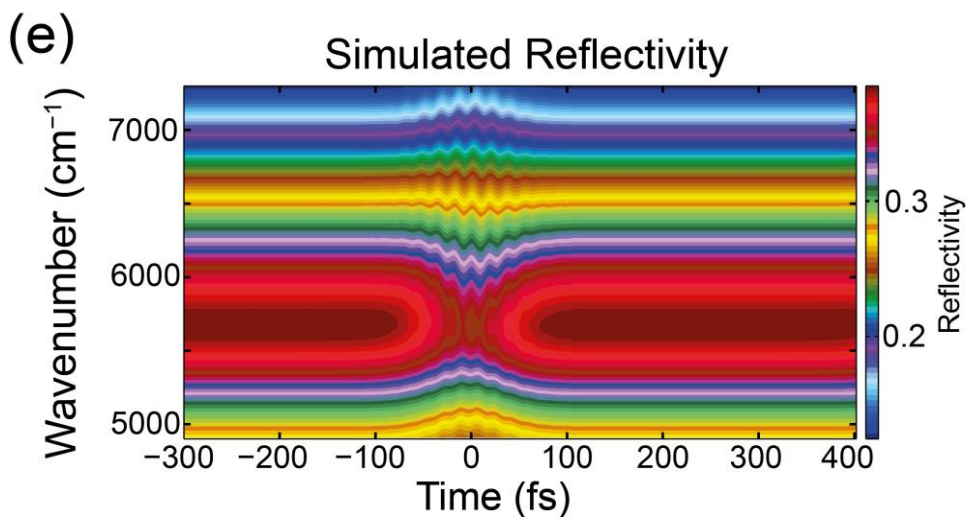
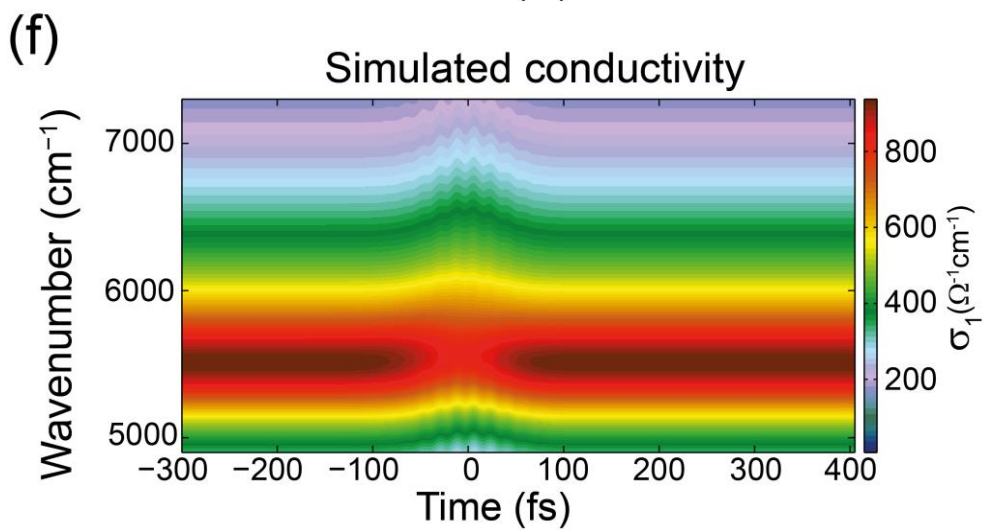

Figure 5: (a) and (c) Time dependence of effective Hubbard correlation term $U/t$ and $V/t$ respectively extracted from fits. (b) and (d) Assumed $U/t$ and $V/t$ variation respectively over time for numerical simulation, similar to one obtained experimentally. (e) Simulated reflectivity obtained via two-time current-current correlation function calculations of an effective strong-coupling model. (f) Corresponding simulated conductivity from the calculations.

## Supplementary Information

### A. Steady state theory fit of the optical conductivity

Here we expand our discussion on the fitting scheme used to extract contributions from holon-doublon (HD) and particle-hole (PH) continuum in the optical conductivity spectra, using a procedure developed in Ref. [27]. In the limit of large $U \gg t, V$ it is possible to analyze the extended Hubbard model by means of a $1/U$ expansion [28]. If we ignore corrections of order $t/U$ then electron transfers are limited to those which conserve the number of doublons and in the ground state all sites are singly occupied. In this limit spin and charge dynamics decouple, leading to spin-charge separation, and a simple upper and lower Hubbard band picture emerges for the charges. The spin-dependence of the optical conductivity then enters only via a momentum dependence ground-state spin correlation $g_q$. This is further simplified by using the so-called "no-recoil approximation" where the dominant contributions to the conductivity arise from $q = 0$ and $q = \pi$ transitions [28], where $g_0 = 2.65$ and $g_\pi = 0.05$. The reduced optical conductivity is then given by

$$\omega \sigma_1(\omega) = \pi g_\pi t^2 e^2 \delta(\omega - \omega_2) + g_0 t^2 e^2 \left\{ \Theta(V - 2t) \pi \left(1 - \frac{4t^2}{V^2}\right) \delta(\omega - \omega_1) + \Theta(4t - |\omega - U|) \frac{2t\sqrt{1 - (\omega - U/4t)^2}}{V(\omega - \omega_1)} \right\} \quad (A1),$$

where $\Theta(x)$ is the Heaviside step function, $\omega_1 = U - V - 4t^2/V$ is the exciton energy and $\omega_2 = U - V$. The two $\delta$ peaks correspond to two different Mott-Hubbard excitons, with the $\Theta(V - 2t)$ factor expressing that the $\omega_1$ exciton exists only when $V > 2t$. Owing to the small value of $g_\pi$ the $\omega_2$ exciton can be ignored in our analysis due to its negligible weight. The final term is a semi-elliptic contribution from the particle-hole continuum which exists only for $|\omega - U| \leq 4t$.

This strong coupling result for the optical conductivity, Eq. (A1), can be reproduced by an effective single-particle model in a semi-infinite single-band tight-binding chain of length $L$ with open boundary conditions. As shown in Fig. 1 this effective model has a zero potential energy for the ground state configuration $|g\rangle$, an energy $U - V$ for the adjacent holon-doublon configuration $|0\rangle$ and all the more distant holon-doublon configurations $|l\rangle$, where $l = 1,2,3,\cdots L$ have an energy $U$. The hopping between neighbouring holon-doublon configurations, e.g. $|l - 1\rangle$ and $|l\rangle$, is included, but no hopping occurs between $|g\rangle$ and $|0\rangle$, consistent with the strong-coupling limit of the extended Hubbard model where holon and doublon populations are conserved by coherent evolution. The total Hamiltonian reads

$$H = (U - V)|0\rangle\langle 0| + U \sum_{l=1}^{L} |1\rangle\langle 1| - 2t \sum_{l=0}^{L-1} (|l\rangle\langle l + 1| + |l + 1\rangle\langle l|). \text{(A2)}$$

The current operator in this picture can be taken as $J \propto |g\rangle\langle 0| + |0\rangle\langle g|$ which mimic the optical excitation of $|g\rangle$ through the creation of an adjacent (bound) holon-doublon configuration $|0\rangle$. The optical conductivity of the model $\sigma_1(z) = \langle g|J(z - H)^{-1}J|g\rangle$ is then equivalent to the Green function $G(z) = \langle 0|(z - H)^{-1}|0\rangle$, with complex frequency $z$. In the limit $L \to \infty$ this is found to be

$$G(z) = \frac{2}{\sqrt{z - U + 2V \pm \sqrt{(z - U)^2 - (4t)^2}}}$$

Expanding the imaginary part of $\omega\, G(\omega)$ and extracting the residue of the poles reproduces the dominant $\omega_1$ Mott-Hubbard exciton $\delta$ peak and the particle-hole continuum found from the strong-coupling result. In other words, apart from the spin-averaging numerical factor $g_0$, the same spectral features of the extended Hubbard model in the limit $U \gg V, t$ are in fact captured by this much simpler effective model. Note also that it is crucial for this effective model to have a hopping amplitude $2t$ so the particle-hole continuum has a bandwidth of $8t$. To allow comparison to the experiment, the sharp features arising from the $\delta$ and $\Theta$ functions were convolved with a Lorentzian inducing a broadening $\eta \sim 2t$. An example of the results of this procedure for the optical conductivity are shown in Fig.2 (a), where we individually show the exciton and particle-hole continuum contributions, along with the resulting fit of the total spectrum.

The subsequent best-fit, performed for each time step, allowed us to extract the effective on-site correlation term *U/t* and effective inter-site interaction term *V/t*. From the experimental results, we found that *U/t* shows overall reduction as a function of pump-probe delay time superimposed by 2Ω oscillation on top. Whereas, *V/t* shows only its gross reduction without the oscillatory character (see Fig.5(c) of the main text). The crucial 2$\Omega$ oscillations of effective correlation term *U/t* underlines the ability of selective molecular vibrational perturbation to couple to on-site interaction term in non-linear manner. The exciton oscillates back and forth in energy about its peak position as evidenced in Fig.2(b). Additionally, the continuum shows overall red shift again pointing towards the overall reduction of U/t (see Fig.2(c)).

The excellent agreement of the experimental observations (Figs. 4a and 4b in the main text) and the calculated optical properties (Figs. 5e and 5f of the main text) emphasizes the importance of the *U/t* modulation. On the contrary, assuming the opposite behaviour with a non oscillatory reduction of *U/t* and an superimposed oscillation on the *V/t* ratio (see Fig. 3a and 3b), the calculated reflectivity and optical conductivity (Fig. 3c and 3d) are substantially different from the experimental observations. Most prominently the reduction of the peak of the excitation and the reshaping of the charge transfer band are not reproduced. That observation strengthens our picture that only the effective onsite-correlations *U/t* get periodically modulated by the vibrational excitation due to the direct modulation of the molecular orbital wavefunction, and no periodic modulation of the nearest neighbor interaction takes place. This is also consistent with the fact that only *U/t* is the only energy scale sufficiently larger than the vibrational frequency to follow its oscillations in an essentially adiabatic way.

**B. Two-time correlation function**

As described in the main text, we took into account the influence of vibrations by considering the oscillators as having a position coordinate described by a scalar $q_j(\tau) = Q_0 [\sin(\Omega_{IR} \tau + \phi)]$, where $Q_0$ and $\phi$ are the maximum displacement and the phase, respectively, identical for all oscillators. This was shown to have the effect of making the $U$ interaction time-dependent.

To reproduce the finite duration of the pulse, we superimposed the sinusoidal oscillation of $U$ on a gaussian envelope. The width and amplitude of the Gaussian is extracted from the fit of Fig. 5a of the main text. Similarly, also the $V$ is assumed to decrease, but without periodic modulation. Overall,

$$U(\tau) = U\{1 - A_U e^{-(\tau-\tau_P)^2/T_P^2}[1 - P_0 q_{IR}^2(\tau)]\}, \qquad V(\tau) = V\{1 - A_V e^{-(\tau-\tau_P)^2/T_P^2}\}.$$

Here, $A_U$ and $A_V$ quantify the overall reduction of $U$ and $V$ ensuing the molecular vibration.

The central quantity to compute is the unequal time current-current correlation function of the initial state, i.e. $\chi_{JJ}(\tau,\tau') = \langle 0|\hat{U}^\dagger(\tau+\tau')\hat{J}\hat{U}(\tau+\tau')\hat{U}^\dagger(\tau')\hat{J}\hat{U}(\tau')|0\rangle$. To this end, we used the effective model, Eq.(A2), as Hamiltonian that governs the time evolution in $\hat{U}$. Let us recall that $|0\rangle$ represents all the configurations with only singly occupied sites, but no specific spin order due to being at room temperature. The current operator $\hat{J}$, in this picture, creates a particle at "site" $|2\rangle$, with energy $U - V$, i.e., the energy to create a nearest-neighbor holon-doublon pair. The explicit time-dependence of $U$ and $V$ mimics the effects of molecular vibrations. Finally, by means of Kramers-Kronig transformation, we obtain the results shown in Fig. 5(e) of the main text.

Figure 1: Effective model describing holon-doublon dynamics in the strong-coupling limit. The state $|g\rangle$ is the ground state containing no holon-doublon states, whereas the state $|0\rangle$ represents an adjacent holon-doublon pair. The remaining states $|l\rangle$ represent the holon and doublon being separated by $l$ sites. In the limit $L \to \infty$ these unbound states form the particle-hole continuum.

Figure 2: (a) Normalized reduced optical conductivity (green) with model fit (dashed black), together with contributions from exciton (red) and continuum (blue) (c) and (d) Normalized time-dependent behavior of the exciton and particle-hole continuum extracted using the fitting procedure detailed in the main text.

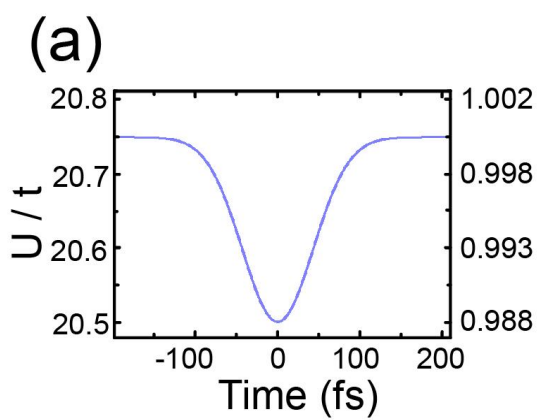
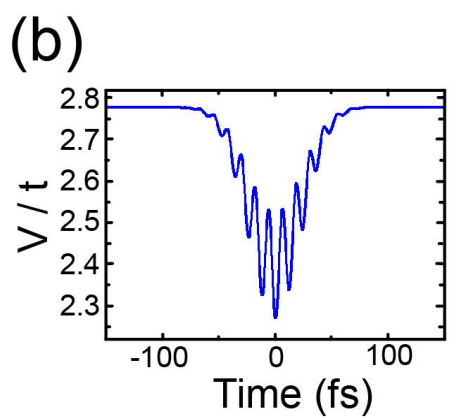
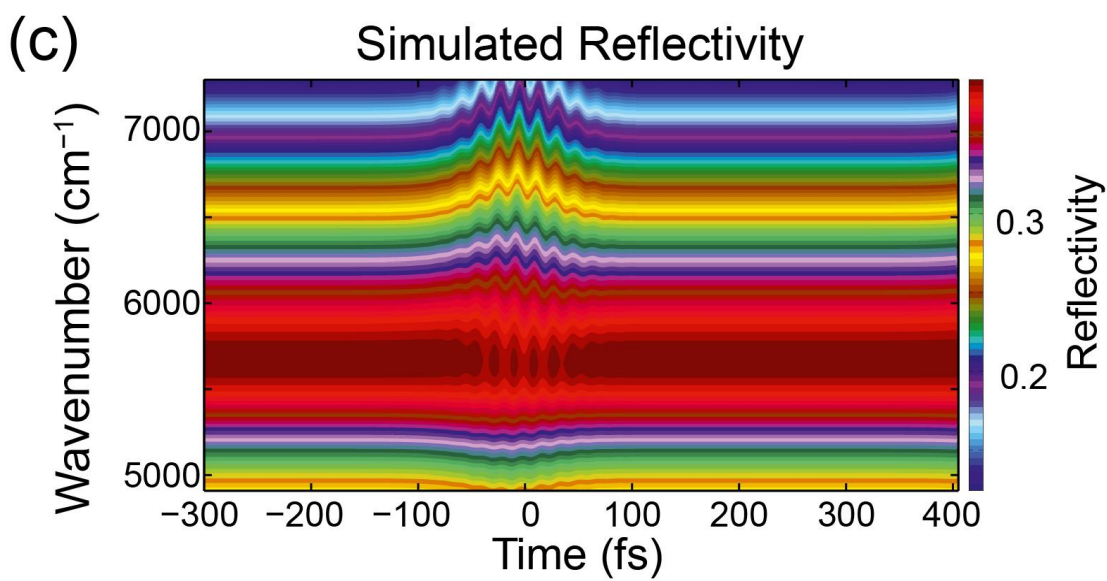
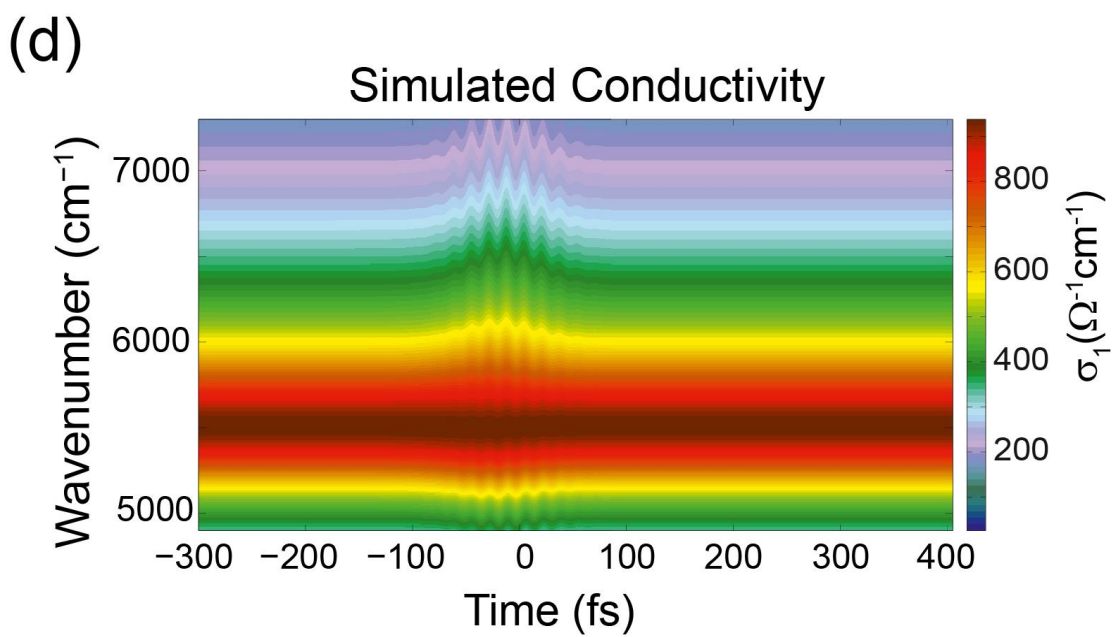

Figure 3: (a) and (b) Assumed variation of $U/t$ and $V/t$ respectively over time for numerical simulation, with fast oscillations superimposed on $V/t$ instead of $U/t$ (c) Simulated reflectivity obtained via two-time current-current correlation function calculations of an effective strong-coupling model (d) Corresponding simulated conductivity from the calculations.